\documentclass{icrc2009}

\usepackage{graphicx}   % for including figures
\usepackage{caption}    % for captions
\usepackage[font=footnotesize]{subfig} % subfig.sty for a double column floating figure using two subfigures
\usepackage{fixltx2e}
\usepackage{url}

\newcommand{\shorttitle}[1]%
{\markboth{Proceedings of the 31\MakeLowercase{$^{st}$} ICRC, {\L}\'{o}d\'{z} 2009}{#1} }
\newcommand{\etal}{\MakeLowercase{\textit{et al. }}} % "et al."

\begin{document}
\title{ A parameterisation of the flux and energy spectrum of single and multiple muons in deep water/ice}

\author{\IEEEauthorblockN{ M. Bazzotti\IEEEauthorrefmark{1},
S. Biagi \IEEEauthorrefmark{1} \IEEEauthorrefmark{2}, 
G. Carminati\IEEEauthorrefmark{1}, \IEEEauthorrefmark{2},
S. Cecchini\IEEEauthorrefmark{2} \IEEEauthorrefmark{3}, 
T. Chiarusi\IEEEauthorrefmark{2},\\ 
G. Giacomelli\IEEEauthorrefmark{1} \IEEEauthorrefmark{2}, 
A. Margiotta\IEEEauthorrefmark{1} \IEEEauthorrefmark{2}, 
M. Sioli\IEEEauthorrefmark{1} \IEEEauthorrefmark{2}, 
M. Spurio\IEEEauthorrefmark{1} \IEEEauthorrefmark{2} }
                            \\
\IEEEauthorblockA{\IEEEauthorrefmark{1} Department of Physics, University of Bologna, Viale Berti Pichat 6/2, I-40127 }
\IEEEauthorblockA{\IEEEauthorrefmark{2} INFN, Sezione di Bologna, Viale Berti Pichat 6/2, I-40127}
\IEEEauthorblockA{\IEEEauthorrefmark{3} INAF-IASF, Via Gobetti 101 Bologna, I-40129}
 }

% please write the preseter's name and short title (3-4 words maximum)
%    which will appear at the header of the even pages.
\shorttitle{M. Spurio \etal Muon flux parameterization}
\maketitle

\begin{abstract}
In this paper parametric formulas are presented to evaluate the flux of atmospheric muons in the range of vertical depth between 1.5 to 5 km of water equivalent (km w.e.)  and up to 85$^o$ for the zenith angle. We take into account their arrival in bundles with different muon multiplicities.  The energy of muons inside bundles is then computed considering the muon distance from the bundle axis. 

This parameterisation relies on a full Monte Carlo simulation of primary Cosmic Ray (CR) interactions, shower propagation in the atmosphere and muon transport in deep water \cite{para_app}. The primary CR flux and interaction models, in the  range in which they can produce muons which may  reach $\sim$1.5 km w.e., suffer from large experimental uncertainties. We used a primary CR flux and an interaction model able to correctly reproduce the flux, the multiplicity distribution, the spatial distance between muons as measured  by the underground MACRO experiment. 
\end{abstract}

 
%%%%%%%%%%%%%%%%%%%%%%%%%%%%
\section{Introduction}
%%%%%%%%%%%%%%%%%%%%%%%%%%%%
  
 The search for neutrinos of astrophysical origin with a km$^3$ scale neutrino telescopes is one of the most important challenges for high energy physics and astrophysics. Atmospheric muons represent the most abundant signal in a neutrino telescope, and are useful for calibration purposes and to check its response to the passage of charged particles. On the other hand, atmospheric muons may represent a background source because they may mimic high energy neutrino interactions; in particular, multiple muons in bundles seem to be particularly dangerous. 
A full Monte Carlo (MC) simulation of atmospheric showers can accurately reproduce the main features of high energy muons reaching a neutrino telescope, at a price of a huge amount of CPU time. Parameterisations of the underground/water atmospheric muon flux are available in the literature \cite{okada,kbs}; they allow an easy comparison of their predictions with experimental results. However, all of them do not take into account the simultaneous arrival of muons in bundles. 

The first author \cite{okada} assumes as muon integral energy spectrum at sea level the empirical formula given in \cite{miyake}. The author obtains an expression for the muon flux at depth $>$1 km w.e. and energy from few GeV up to 10 TeV through a computation of the propagation of the particles in sea water. 

%%%%%%%%%%%%%%%
 \begin{figure}[!t]
  \centering
  \includegraphics[width=3.2in]{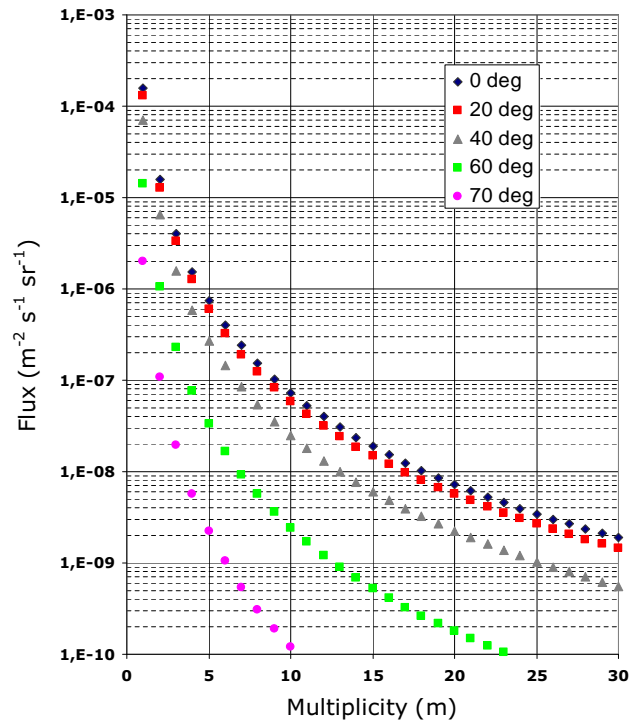}
  \caption{Multiplicity distribution of muons in bundles evaluated with (\ref{eq:eq1}) at the depth of 3 km w.e. and for 5 different values of the zenith angle. The computation assumes here a muon energy threshold of 1 GeV.}
  \label{multi}
 \end{figure}
 %%%%%%%%%%%%%%

The second calculation \cite{kbs}   starts from an analytical expression of the muon spectrum at sea level \cite{gaisser} and  of the muon propagation in rock or water. 
The result is an analytical expression that gives the integral muon flux in the range 1 GeV -10 TeV for depths in the range 1.5-16 km w. e.. The integral flux depends from 5 free parameters, which must be tuned in the expression of the sea level spectrum. 

In both cases no distinction is made between "single" muons and "multiple" muons (or muon bundles).  These fluxes are all-inclusive and contain a sort of  summation over the multiplicity distribution, which actually changes with depth and zenith angle. 

Underwater/ice telescopes \cite{icecube,antares,km3net} have a coarser spatial resolution with respect to underground detectors like MACRO. In particular, they cannot measure the muon multiplicity of atmospheric muon bundles. In addition, their trigger and reconstruction efficiencies strongly depend from the bundle multiplicity.   
The present parameterisation (full details in \cite{para_app}) reproduces the flux of atmospheric muons as a function of their multiplicity, vertical depth and zenith angle (see Fig. \ref{multi}). The energy spectrum is given as a function of the muon distance from the axis shower.  

From the parametric formulas, an event generator (MUPAGE, \cite{mupage}) was derived. It is well suited to be interfaced with MC simulations of atmospheric muons of underwater/ice experiments. Details and some application of this generator are presented in \cite{giada}.

%%%%%%%%%%%%%%%%%%%%%%%%%%%
\section{From full MC to parametric formulas}
%%%%%%%%%%%%%%%%%%%%%%%%%%%
The present parameterisation of the multiple muon flux and energy spectra relies on a full MC simulation of the primary CR flux, interactions and shower propagation in the atmosphere. 
The latest version of the HEMAS  code was used (HEMAS-DPM \cite{hemasdpm}). The reliability of the code is restricted to secondary particles with energies above $E_T=$ 500 GeV. Muons below  $E_T$ (at sea level) are not followed through water because their survival probability  for depths $> 1.5$ km w.e. is very small. The so called {\it prompt muons} (from the decay of charmed mesons)  are not included. 

HEMAS was preferred to other codes since it was deeply used and cross-checked with the results of the underground MACRO experiment. Also the input primary CR spectrum was a phenomenological model obtained by MACRO \cite{macro-comp2}. 
Both CR composition and some of the hadronic interaction features in HEMAS were tuned \cite{tesi-max} by studying the experimental muon multiplicity distribution \cite{macro-comp1}, the underground muon pair distance distribution (the so-called decoherence distribution \cite{macro-deco}) and the average energy of single and double muons \cite{macro_trd}. These measurements are particularly important because constrain some quantities extrapolated in the MC outside the kinematics region measured at accelerators, as the charged multiplicity distributions and the transverse momentum of muon parent mesons at TeV energies.

%%%%%%%%%%%%%%%
 \begin{figure}[!t]
  \centering
  \includegraphics[width=3.1in]{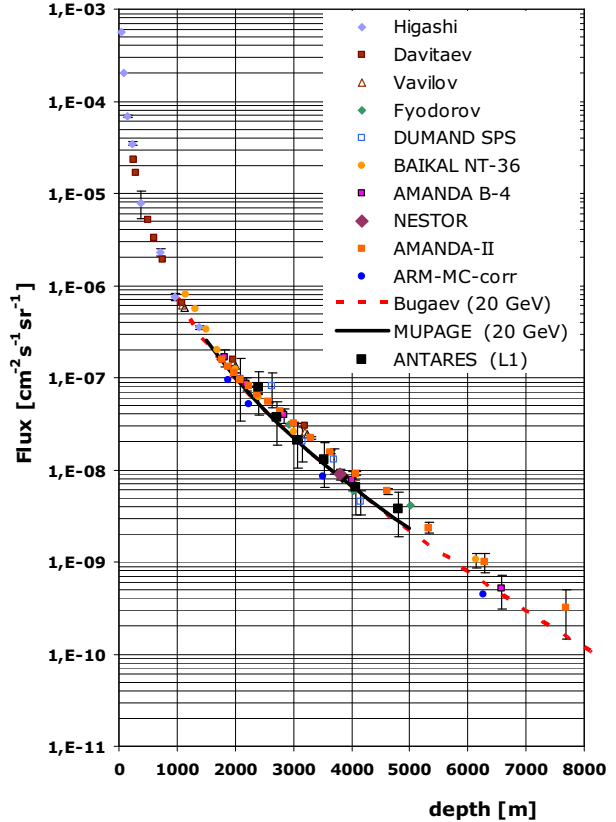}
  \caption{Muon vertical intensity versus  depth of water or ice. Comparisons are shown among experimental points (reference in \cite{kbs}), the ANTARES line 1 measurement \cite{line1}, our parameterisation (full line) and the one of \cite{kbs} (dashed line). Both computations considered only muons with energy $>20$ GeV at a given depth.  The ANTARES measurement with 5 lines is reported in \cite{marco}.}
  \label{dir}
 \end{figure}
 %%%%%%%%%%%%%%

The muons in the atmospheric shower reaching the sea-level (flat surface) were  propagated down to 5 km of water using  MUSIC \cite{music}, a 3D muon propagation code which uses recent and accurate cross sections of the muon interactions with matter. 

All the muon kinematics information from $h=$1.5 km w.e. down to $h=$5.0 km w.e. were stored in ROOT-format files.
We consider the vertical depth $h$ as one of the main parameters, as the zenith angle $\theta$. The flux is computed up to $\theta=85^o$.
% so the maximum value of $X=h/cos\theta\simeq 50$ km w.e. . 
From these files the parameters of the analytic description of the muon flux and of the energy spectra were obtained with a fitting procedure. The energy of the muons in bundles depends (at a fixed $h$ and $\theta$) on the bundle multiplicity $m$ and on the distance $R$ of each muon from the bundle axis.

The flux (units: $m^{-2} s^{-1} sr^{-1}$) of muon bundles with multiplicity $m$ (see Fig. \ref{multi}) is obtained as a function of vertical depth $h$ and zenith angle $\theta$ as:
\begin{equation}
\Phi(m;h,\theta)= {K(h,\theta) \over m^{\nu(h,\theta)}} 
\label{eq:eq1}
\end{equation}
The flux of bundles of increasing multiplicity $m$ decreases with increasing vertical depth and zenith angle. The parameters $ K(h,\theta)$ and $\nu(h,\theta)$ in (\ref{eq:eq1}) depend from 9 constants, which are reported in \cite{para_app}.
From (\ref{eq:eq1}), the  depth-intensity relation for vertical muons ($\theta =0^o$) can be calculated, and it is presented in Fig. \ref{dir}.

The energy spectrum of muons is described \cite{gaisser} by:
\begin{equation}
{ dN \over d (log_{10}E_\mu) } = G E_\mu e^{\beta X (1-\gamma)} [E_\mu + \epsilon (1-e^{-\beta X})]^{- \gamma}    
\label{eq:spectrum} 
\end{equation}
where $\gamma$ is the spectral index of the primary CRs and $\epsilon= \alpha /\beta$ is defined from the parameters of the muon energy loss formula $-\langle{dE(E_\mu) \over dX}\rangle = \alpha + \beta E_\mu $. 
In this work, $\gamma$ and $\epsilon$ were used as free fit parameters and $\beta$ was fixed. The constant $G=G(\gamma,\epsilon)$ represents a normalization factor, in order that the integral over the muon energy spectrum (\ref{eq:spectrum})
from 1 GeV  to 500 TeV is equal to 1. In the case on single muon events (bundles with $m=1$) the $\gamma=\gamma(h) , \ \epsilon=\epsilon(h,\theta)$ parameters depend on 6 constants. 

\begin{figure}[ht]
  \includegraphics[width=3.2in]{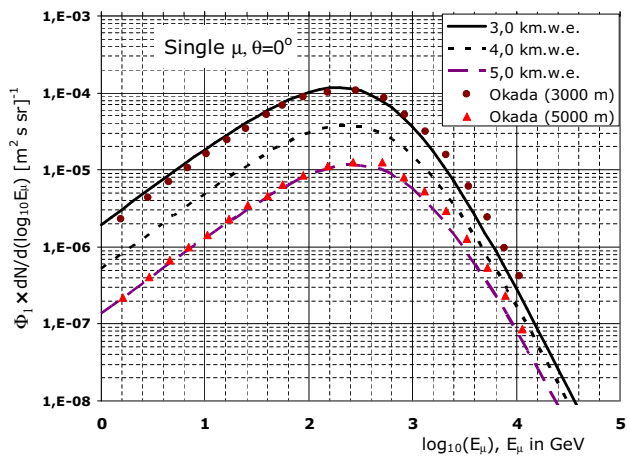}
\caption{Expected energy spectrum of single muons at $\theta=0$ and three different depths. The points represent the predictions (normalized to the same number of events) extracted from Fig. 3 of \cite{okada} (see text). }
\label{okada} 
\end{figure}

%%%% NUOVO
Fig. \ref{okada} shows the evaluated energy distribution of single muons at $\theta=0$ and at various depths ($h$= 3, 4 and 5 km w.e.). It is evaluated by multiplying the normalized energy distribution (\ref{eq:spectrum}) with the flux (\ref{eq:eq1}) of single muons  $\Phi_1=\Phi(m=1,h,\theta=0)$. For instance, at $h=3$ km w.e., $\Phi_1=1.6\times 10^{-4} \ (m^2\ s\ sr)^{-1}$. The total muon flux at the same $h$ and $\theta$ is $\Phi_T=\sum_{m} m \Phi(m,h,\theta) = 2.2\times 10^{-4} \ (m^2\ s\ sr)^{-1}$. For comparison, the result of the computation of \cite{okada} is reported. In this case, the total muon flux at $h=3$ km w.e., $\theta=0$ is 2.1$\times 10^{-4} \ (m^2\ s\ sr)^{-1}$. In order  to compare with the distribution of our single muons, we normalized in Fig. \ref{okada} the prediction of  \cite{okada} to the same number of muons.
%%%%%%%%%%%%%%%%%%%
%%%%%%%%%%%%%%%%%%%%%
%%%%%%%%%

The situation is more complicated for multiple muons. Due to the muon production kinematics, the muon energy depends from their distance with respect to the axis bundle. The description of the muon lateral distance $R$ from the axis  bundles is thus the preliminary step to evaluate the muon energy distribution in a bundle.
$R$ (in the plane orthogonal to the axis) was extracted from a distribution of the form: 
\begin{equation}
{dN \over dR} = C { R\over (R+R_0)^\alpha }
\label{eq:radial} 
\end{equation}
The parameters $R_o=R_o(h,m,\theta)$ and $\alpha=\alpha(h,m)$ depend from 9 constants. 
%%%%%%%%%%%%%%%%%%%
\begin{figure}[ht]
  \includegraphics[width=3.5in]{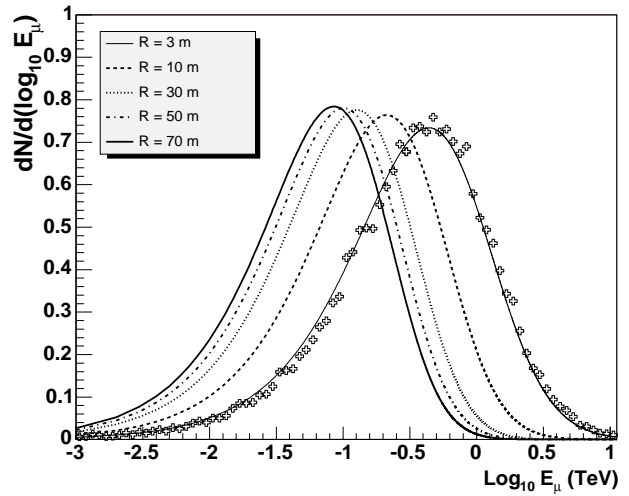}
\caption{ Differential energy spectra of double muons ($m=2$), at five different distances  from the bundle axis: $R=3, 10, 30, 50$ and $70 \ m$. The lines were computed with (\ref{eq:spectrum}) using the 15 constants reported in Tab. 6 of \cite{para_app}, assuming  $\theta=0$ and $h=3.5$ km  w.e. = 3500 hg cm$^{-2}$. The marker points superimposed for references to the line $R=$3 m were obtained with the full MC simulation. 504 series of points were used by the fitting procedure to obtain the 15 constants.}
\label{emum} 
\end{figure}
%%%%%%%%%%%%%%%%%%%%%

The energy spectrum of muons arriving in bundles has the same general form as for single muons (\ref{eq:spectrum}). In the case of multiple muons, the analytic description of the parameters $\gamma=\gamma(h,R,m)$ and $\epsilon=\epsilon(h,R,\theta)$ depend from 15 constants. An example of the energy distribution of double muons ($m=2$) is shown in Fig.  \ref{emum}.

%%%%%%%%%%%%%%%%%%%%%%%%%%%%%%
\section{Discussion of the results}
%%%%%%%%%%%%%%%%%%%%%%%%%%%%%%
The present parameterisation allows to evaluate not only the \textit{total muon flux}, but also the total number of \textit{muon bundles} in deep detectors. The systematic uncertainties on the predictions are related to the all CR particle spectrum,  chemical composition and interaction model. 
It should be noted that many experiments, like in neutrino telescopes,  cannot measure the muon bundle multiplicity.  The relevant quantity in calculating the total muon flux is the number of $nucleons$ impinging at the top of the atmosphere.  For the number of muon bundles, it seems more relevant the number of $nuclei$ at the top of the atmosphere. The relative contribution to the total number changes significantly from one primary CR model to the other.

%% Primary Flux
The knowledge of the all-particle spectrum is limited by the gap from "direct" and "indirect" observations \cite{horandel}. The overlap will be crucial in establishing a reference point of the latter and for modelling the origin of CRs at energies of the knee and beyond. At the energies of interest for neutrino telescopes, the primary CR composition deduced from experimental data shows a strong dependence on the interaction models adopted (see i.e. \cite{kaskade}). Deviations from one parameterisation to another vary from 10\% to 100\% in absolute value and, particularly, the position of the knee differs from one model to the other. 

%% Interaction model
The L3+C data up to 2 TeV show that the muon flux predicted using different interaction models can differ by up to 30\%. At higher energies the situation is worse. An overall uncertainty on the \textit{total muon flux} and on the number of \textit{muon bundles} vary from 30\% to more than 100\% at different depths and zenith angles.
%%%%%
The measurements of the deep-intensity relation constrain the total vertical flux of muons  at different depths (Fig. \ref{dir}) by not better than 30\%. 

The approach of this work was to choose a combina\-tion of the primary CR flux, CR composition and interaction model which reproduces (at the level of $\sim$30\%) the MACRO data (depth: $2000-5000$ hg cm$^{-2}$, $\theta<60^o$).

\begin{figure}[ht]
  \includegraphics[width=3.2in]{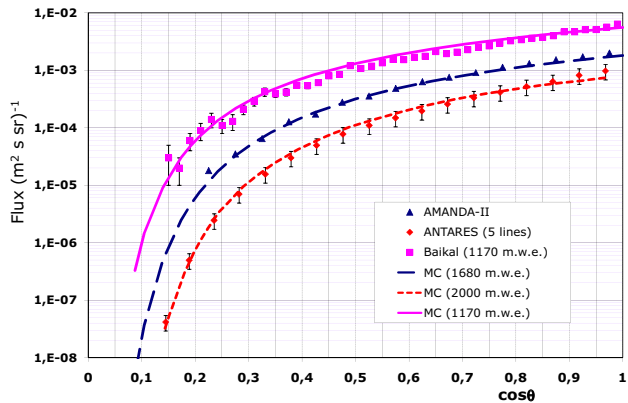}
\caption{Muon flux as a function of the $cos\theta$ as measured by ANTARES \cite{marco}, AMANDA-II \cite{desiati}, and Baikal \cite{baideep} at three different depths. The results of our computation for a muon energy threshold of 20 GeV at the three depths is superimposed as full lines.}
\label{cosze} 
\end{figure}

Fig. \ref{cosze} shows the comparisons of the zenith distribution evaluated at a fixed depth using (\ref{eq:eq1}) and some experimental data.
%%%%%%%%%%%%%%%%%%%
%%%%%%%%%%%%%%%%%%%%%

The comparisons demonstrate that the analytical solution provides a good representation of the experimental data so far collected. This is true also in the case of the Baikal experiment, at a depth nominally outside the range of validity of our parameterization.  One has to consider that in many cases the analyses are still approximate and not all the experiments have declared their systematic uncertainties. 
%The measurement of the angular flux of HE muons ($E_\mu>10$ TeV) is mandatory way for identifying the prompt component of atmospheric neutrinos (isotropically distributed) which constitute a background for neutrino telescope.

\begin{figure}[!t]
  \centering
  \includegraphics[width=3.0in]{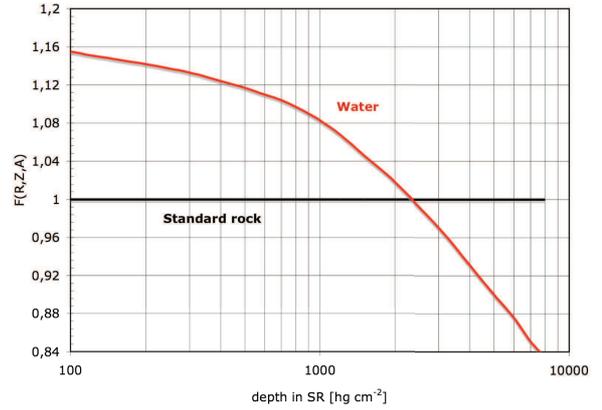}
  \caption{Factor by which depths in water must be multiplied to convert to standard rock. For example, 1.08 km of water are equivalent to 1000 hg cm$^{-2}$ of standard rock \cite{wri}. }
  \label{srwater}
 \end{figure}

The parametric formulas can be used also for underground experiments, provided that the correct conversion to equivalent standard rock is made. The parameters of the muon energy losses $-\langle{dE(E_\mu) \over dX}\rangle = \alpha + \beta E_\mu $ depend from the medium: at small depths, the water is a more effective absorber (per g/cm$^2$) because the ionization term $\alpha$ is greater than that for rock. At larger depths ($> 1$ km) the  $\beta$ terms begin to dominate and made rock a more effective absorber. This effect is illustrated in Fig.   \ref{srwater}.

%%%%%%%%%%%%%%%%%%
%%%%%%%%%%%%%%%
 %%%%%%%%%%%%%%

%%%%%%%%%%%%%%%%%%%%%%%%%%%%%%
 \section{Conclusions}
%%%%%%%%%%%%%%%%%%%%%%%%%%%%%%
A new parameterisation of the atmospheric muon flux under a homogeneous medium of water/ice, considering the simultaneous arrival of muons in bundles, is presented.   It allows to evaluate the \textit{total muon flux} and the total number of \textit{muon bundles} in deep detectors. Flux,  zenith angle distributions and average energies obtained with these formulas are in good agreement with those predictions by other parameterisations and with  experimental measurements. This parameterisation is well suited  as a fast generator to simulate the atmospheric muon flux in underwater/ice neutrino telescopes, particularly for the neutrino background simulation.


\begin{thebibliography}{99}
\bibitem{para_app} Y. Becherini et al., Astropart. Phys. 25 (2006) 1 
\bibitem{okada} A. Okada, Astropart. Phys. 2 (1994) 393; ICRR-Rep-319-94-14
\bibitem{kbs} S. Klimushin, E. Bugaev, I. Sokalski, Phys. Rev D64 (2001) 
14016; S. Klimushin, E. Bugaev,I. Sokalski hep-ph/0106009
\bibitem{miyake} S. Miyake Proc. 13th ICRC (Denver, 1973) 5, 3638
\bibitem{gaisser} T.K. Gaisser, Cosmic rays and particle physics, Cambridge Univ. Press (1990)
\bibitem{icecube} C. Portello-Roucelle (IceCube coll.) arXiv:0805.3546 [astro-ph]
\bibitem{antares} M. Spurio (ANTARES coll.) arXiv:0904.3836 [astro-ph]
\bibitem{km3net} see: http://www.km3net.org/ 
\bibitem{mupage} G. Carminati et al., Comput. Phys. Commun.179(2008) 915. 
\bibitem{giada} G. Carminati et al., this ICRC09 conference.
\bibitem{hemasdpm} G.~Battistoni {\em et al.}, Astropart. Phys. {\bf 3}
(1995) 157; INFN/AE-99/07 (1999).
\bibitem{macro-comp2} M.~Ambrosio {et al.} (MACRO Collaboration), Phys. Rev. {\bf D56} (1997) 1418.
\bibitem{tesi-max} M.~Sioli, {\em A new approach to the study of high 
energy muon bundles with the MACRO detector at Gran Sasso}, Ph.D Thesis,
Bologna University, hep-ex/0209029
\bibitem{macro-comp1} M.~Ambrosio {et al.} (MACRO Collaboration), 
Phys. Rev. {\bf D56} (1997) 1407.
\bibitem{macro-deco} M.Ambrosio { et al.} (MACRO Collaboration ), Phys. Rev. {\bf D60} (1999) 032001.
\bibitem{macro_trd} M. Ambrosio et al. (MACRO Collaboration), Astrop. Phys. {\bf 19} (2003) 313.
\bibitem{music} P. Antonioli et al. , Astrop. Phys. 7(1997) 357.
\bibitem{line1} M. Ageron et al. (ANTARES Coll.) Astrop. Phys. 31(2009) 277.
\bibitem{marco} M. Bazzotti for the ANTARES collaboration. This ICRC09 conference. 

\bibitem{desiati}  P. Desiati, K. Bland et al. (AMANDA Collaboration), {\em Response of AMANDA-II to Cosmic Ray Muons}, 28th ICRC, Tsukuba, Japan, HE 2.3, 1373-1376
\bibitem{baideep} I.A. Belolaptikov et al. Astrop. Phys. 7(1997)263.
\bibitem{horandel} J. Horandel, Astrop. Phys. 19 (2003) 193.
\bibitem{kaskade} T. Antoni et al. , Astrop. Phys. 25(2005)1.
\bibitem{wri} A.G. Wright, Proc. 13th ICRC (Denver, 1973) 3, 1709


  \end{thebibliography}
\end{document}